\documentclass[a4paper,11pt]{article}
\pdfoutput=1 

\usepackage{jheppub} 
\usepackage{color}
\usepackage[T1]{fontenc} 

\title{Competition between the s-wave and p-wave superconductivity phases in a holographic model} \vskip 2cm \vskip 2cm
\author[a,b,c]{Zhang-Yu Nie,}
\author[b]{Rong-Gen Cai,}
\author[b,d]{Xin Gao,}
\author[a,d]{Hui Zeng}

\affiliation[a]{Kunming University of Science and Technology, \\Kunming 650500, China}
\affiliation[b]{State Key Laboratory of Theoretical Physics, Institute of Theoretical Physics, Chinese Academy of Sciences,
\\P.O.Box 2735, Beijing 100190, China}
\affiliation[c]{INPAC, Department of Physics, and Shanghai Key Laboratory of Particle Physics and Cosmology, Shanghai Jiao Tong University,
\\Shanghai 200240, China}
\affiliation[d]{Max-Planck-Institut f\"{u}r Physik (Werner-Heisenberg-Institut),\\
F\"{o}hringer Ring 6, 80805 M\"{u}nchen, Germany}

\emailAdd{niezy@kmust.edu.cn}
\emailAdd{cairg@itp.ac.cn}
\emailAdd{gaoxin@mppmu.mpg.de}
\emailAdd{zenghui@kmust.edu.cn}

\abstract{We build a holographic superconductor model with a scalar triplet charged under an SU(2) gauge field in the bulk. In this model, the s-wave and p-wave condensates can be consistently realized. We find that there are totally four phases in this model, namely, the normal phase without any condensate, s-wave phase, p-wave phase and the s+p coexisting phase. By calculating Gibbs free energy, the s+p coexisting phase turns out to be thermodynamically favored once it can appear. The phase diagram with the dimension of the scalar operator and temperature is drawn. The temperature range for the s+p coexisting phase is very narrow, which shows the competition between the s-wave and p-wave orders in the superconductor model.}

\arxivnumber{1309.2204}

\begin{document}
\maketitle
\flushbottom

\section{\bf Introduction }
In the holographic model of superconductors~\cite{Gubser:2008px,Hartnoll:2008vx}, one uses the hair formation near a black hole horizon to mimic the superconductivity phase transition in the condensed matter physics. According to the AdS/CFT correspondence~\cite{Maldacena:1997re,Gubser:1998bc,Witten:1998qj}, the hair on the black hole can be mapped to the expectation value of the condensed operator in the dual field theory. Thus one can use the hair as the order parameter in the holographic superconductivity phase transition. The holographic superconductor model turns out to be quite successful in giving the qualitative features of superconductivity.  The scalar hair in the holographic superconductor model~\cite{Gubser:2008px,Hartnoll:2008vx} is dual to the condensate of scalar operator in the field theory. Therefore it corresponds to an s-wave superconductor model. The holographic superconductor model has also  been generalized to the p-wave case in ref.~\cite{Gubser:2008wv} by using an SU(2) Yang-Mills field in the hulk to realize the condensate of a vector operator. More studies on the holographic p-wave superconductivity can be find in refs.~\cite{Ammon:2009xh,Zeng:2010zn,Cai:2010cv,Zayas:2011dw,Gangopadhyay:2012gx,Roychowdhury:2013aua}.

The superconducting materials may have multi-band structures and can be described by more than one order parameters. There also could be order parameters with different symmetries in one material. For a system with both s-wave and p-wave order parameters, it may be dominated by s-wave phase in some situation and by p-wave phase in another situation. Especially, there could exist both s-wave and p-wave paired electrons and a so-called s+p wave phase in some material~\cite{Goswami:2013}. Then it would be interesting to build a holographic model containing both the s-wave and p-wave phases. One can use this model to study the competition between the s-wave and p-wave superconductivity phases, and furthermore, to study the possibility of the s+p coexisting phase.

Recently, some attentions indeed have been paid to the study on the competition  and coexistence of multi order parameters in the holographic superconductor models. In ref.~\cite{Basu:2010fa} the authors considered the competition effect of two scalar order parameters in the probe limit and found the signal of a coexisting phase where both the two s-wave order parameters have non zero value.  In ref.~\cite{Cai:2013wma}, the authors considered two scalars charged under the same U(1) gauge field with full back reaction of matter fields on the background geometry.  It turns out that the model has a rich phase structure for the competition and coexistence of the two scalar orders. There exist totally five different superconductivity phases and three of them are of the coexistence of the two scalar orders and they are all thermodynamically favored than the phase with only one order parameter.   The authors of ref.~\cite{Huang:2011ac} considered three scalars in the fundamental representation of SO(3) group to model a multi-band superconductor model with more than one s-wave order parameters, while ref.~\cite{Krikun:2012yj} studied an $S^{+/-}$ two band superconductor model by two SU(2) charged bulk scalars. The AC conductivity shows a peak in the mid-infrared which is argued to be related to the inter-band interaction in iron-based superconductors. The authors of ref.~\cite{Musso:2013ija} studied an unbalanced holographic model by two scalars charged under two U(1) gauge fields, which are explained as the electrically superconductivity order and the electrically neutral magnetization respectively.  Other studies concerning the competition of multi order parameters can be found in refs.~\cite{Kim:2013oba,Nitti:2013xaa,Liu:2013yaa,AAJL,DG,DGSW,Wen:2013ufa}. Most of those works concern two orders with the same symmetry, so it would be interesting to study the competition of orders with different symmetry.

In this paper, we build a holographic superconductor model with a scalar triplet charged under an SU(2) gauge field. In this model both the s-wave and p-wave superconductivity phases naturally arise and we find that an s+p coexisting phase can also be realized when the dimension of the scalar operator is in some region. This article is organized as follows. In section~\ref{sect:setup} we set up our model and shortly review the behavior of the s-wave and p-wave superconductivity in this model. In section~\ref{sect:competition}, we study the competition of the s-wave and p-wave condensates by comparing their critical temperature and free energy. We study the s+p coexisting phase in section~\ref{sect:s+p} and find the coexisting phase is thermodynamically favored, compared to the phase with a single order. Our conclusions and discussions are included in section~\ref{sect:conclusion}.

\section{Holographic model of an s+p superconductor}
\label{sect:setup}

\subsection{ The model setup}
To realize the s-wave and p-wave superconductivity in one model, we consider a real scalar triplet charged
 in an SU(2) gauge field in the gravity side. The full action is
\begin{eqnarray}
S  &=&S_G+S_M,\\
S_G&=&\frac{1}{2 \kappa_g ^2}\int d^{d+1}x \sqrt{-g} (R-2\Lambda),\\
S_M&=&\frac{1}{g_c^2}\int d^{d+1}x \sqrt{-g}(-D_\mu \Psi^{a} D^\mu \Psi^a-\frac{1}{4}F^a_{\mu\nu}F^{a\mu\nu}-m^2 \Psi^a\Psi^a). \label{Smatter}
\end{eqnarray}
Here $\Psi^a$ is the SU(2) charged scalar triplet in the vector representation of the SU(2) gauge group, and
\begin{equation}
D_\mu\Psi^a=\partial_\mu \Psi^a+\varepsilon^{abc}A^b_\mu\Psi^c.
\end{equation}
$F^a_{\mu\nu}$ is the gauge field strength, and is given by
\begin{equation}
F^a_{\mu\nu}=\partial_\mu A^a_\nu-\partial_\nu A^a_\mu +\varepsilon^{abc} A^b_\mu A^c_\nu.
\end{equation}
$g_c$ is the Yang-Mills coupling constant as well as the SU(2) charge of $\Psi^a$. We can redefine the fields $A^a_{\mu}$ and $\Psi^a$ to get the standard expression where the coupling $g_c$ appears in the derivative operator $D_\mu$. In this paper, we study the  system in the probe limit. This means that the back reaction of the matter fields on the background geometry is neglected. This limit can be realized consistently by taking the limit $g_c\rightarrow\infty$.

In the probe limit, we choose the background geometry to be a $d+1$ dimensional AdS black brane with metric
\begin{equation}
ds^2=-f(r)dt^2+\frac{1}{f(r)}dr^2+r^2dx^{2}_i.
\end{equation}
$x^i$s are the coordinates of a $d-1$ dimensional Euclidean space.
The function $f(r)$ is
\begin{equation}\label{fEinstein}
f(r)=r^2(1-(\frac{r_h}{r})^d),
\end{equation}
with $r_h$ the horizon radius. Here we have set the AdS radius $L$ to be unity. The temperature of the black brane is related to $r_h$ as
\begin{equation}\label{TemperatureE}
T=\frac{d }{4\pi}r_h.
\end{equation}
This is just the temperature of dual field theory in the AdS boundary.

Let us consider the following ansatz  for the matter fields
\begin{eqnarray}
\Psi^3=\Psi_3(r),\quad A^1_t=\phi(r),\quad A^3_x=\Psi_x(r),
\end{eqnarray}
with all other field components being turned off. In this ansatz, we take $A^1_\mu$ as the electromagnetic U(1) field as in ref.~\cite{Gubser:2008wv}. With this ansatz, the equations of motion  of matter fields in the AdS black brane background read
\begin{eqnarray}\label{eoms}
\phi''+\frac{d-1}{r}\phi' -\Big(\frac{2  \Psi_3^2}{f}+\frac{\Psi_x^2}{r^2 f}\Big)\phi&=&0, \label{eqphi}\\
\Psi_x''+\Big(\frac{d-3}{r}+\frac{f'}{f}\Big)\Psi_x'+\frac{\phi^2}{f^2}\Psi_x&=&0, \label{eqpsix}\\
\Psi_3''+\Big(\frac{d-1}{r}+\frac{f'}{f}\Big)\Psi_3'-\Big(\frac{m^2}{f}-\frac{\phi^2}{f^2}\Big)\Psi_3&=&0. \label{eqpsi3}
\end{eqnarray}
We can see $\Psi^3$ and $\Psi_x$ are not coupled in their equations of motion, and they both coupled to the same U(1) electromagnetic field.

In this model, we can easily realize the p-wave superconductivity by turning off the scalar degrees of freedom. And in the similar way, we can turn on only one scalar component and one gauge field component
to get the s-wave condensation. Looking at the equations of motion, we can see this more clearly. If we turn off $\Psi_x$, one can get the same equations of motion for the holographic s-wave superconductor model in ref.~\cite{Hartnoll:2008vx}. On the order hand, if turn off $\Psi^3$ instead, we obtain the same equations of motion for the holographic p-wave superconductor model in ref.~\cite{Gubser:2008wv}. As the equations of motion for the two cases are respectively the same as the ones in the s-wave or p-wave model, the condensate behavior will be the same as the s-wave or p-wave case individually. As a result, this model can realize  the s-wave and p-wave superconductivity consistently. Here what we can do further is to study the competition between the p-wave phase and s-wave phase by comparing the critical temperature and free energy of different condensations. We can also consider the existence of the coexisting phase of the two orders. Before this we first simply discuss  the s-wave and p-wave condensates in this model.

\subsection{S-wave condensate}
\label{subsect:swave}
The case with only the s-wave condensate can be reached by turning off $\Psi_x$ in eq.~(\ref{eqpsix}). With the remaining two equations, we can find the solutions with nonzero $\Psi_3$. The boundary conditions at the horizon are
\begin{eqnarray}\label{shorizon}
\phi(r)&=& \phi_1(r-r_h)+\mathcal{O}\big((r-r_h)^2\big),\nonumber \\
\Psi_3(r)&=& \Psi_0 + \mathcal{O}(r-r_h).
\end{eqnarray}
Since $\phi(r)$ is the $t$ component of the vector field $A_\mu^1$, so $\phi(r_h)$ is set to zero to avoid the potential divergence of $g^{\mu\nu}A_\mu^1 A_\nu^1$ at the horizon. And eq.~(\ref{eqpsi3}) impose a constraint on the derivative of $\Psi_3(r)$ at the horizon. So the expansions of $\phi(r)$ and $\Psi_3(r)$ have only one free parameter. In other words, the solutions to the equations (\ref{eoms}) and (\ref{eqpsix}) are determined by the two parameters $\phi_1$ and $\Psi_0$.

The expansions of $\phi$ and $\Psi_3$ near the AdS boundary are of the forms
\begin{eqnarray}\label{sboundary}
\phi(r)&=& \mu-\frac{\rho}{r^{d-2}}+..., \nonumber\\
\Psi_3(r)&=& \frac{\Psi_-}{r^{\Delta_-}} + \frac{\Psi_+}{r^{\Delta_+}}+... ~.
\end{eqnarray}
From the AdS/CFT dictionary, $\mu$ and $\rho$ are related to the chemical potential and charge density of the boundary field theory.
$\Psi_-$($\Psi_+$) is the source of a scalar operator with dimension $\Delta_+$($\Delta_-$), and $\Psi_+$($\Psi_-$) is the corresponding expectation value of the operator. The operator dimension is related to the mass of the scalar field as
\begin{equation}
\Delta_\pm=\frac{d\pm \sqrt{d^2+4 m^2}}{2}.
\end{equation}
Here we can see $\Delta_+\geq d/2$ and $\Delta_-\leq d/2$. There is an additional constraint $\Delta_-> (d-2)/2$ if we choose $\Psi_-$ as the expectation value~\cite{Hartnoll:2009sz,McGreevy:2009xe}. In order to get the solution with a spontaneous condensation, we should turn off the source. That means we should enforce an additional boundary condition $\Psi_-=0$($\Psi_+=0$). In this paper, we always choose the scalar dimension to be $\Delta=\Delta_+$ and set $\Psi_-=0$, because the dimension of the p-wave operator is $d-1$ which is in the region of $\Delta_+$ for $d\geq3$ and in this case the behavior of the s-wave phase is  comparable with the p-wave phase.

To solve the equations of motion with the imposed boundary conditions, we use the shooting method. We solve the equations numerically with the initial value $(\phi_1,\Psi_0)$, and then we can get the values of $(\mu,\rho,\Psi_-,\Psi_+)$ for this solution. We shoot the target $\Psi_-=0$ with different solutions parameterized by $(\phi_1,\Psi_0)$, and finally we get a one parameter solution satisfying the condition $\Psi_-=0$.

This one parameter solution then shows the condensate behavior of the scalar operator with dimension $\Delta_+$. Because the equations of motion and the background metric are all the same as those in refs.~\cite{Hartnoll:2008vx,Horowitz:2008bn}, the condensate behavior is also completely the same. Here we have a free parameter $m$ or equivalently $\Delta$, we can tune this parameter to change the condensate behavior of the s-wave phase.

\subsection{P-wave condensate}
\label{subsect:pwave}
In the p-wave case, we instead turn off the field $\Psi_3$ and get the equations of motion for the fields $\phi$ and $\Psi_x$. The equations of motion are quite similar to those in the s-wave case, so the numerical procedure is the same. But we should remember that $\Psi_x$ is a space component of a vector. And because there is no mass term for the vector $A^3$, the p-wave operator has a fixed conformal dimension and the p-wave condensate can not be tuned as in the s-wave case.

In the p-wave case, the expansions of the field $\phi$ near the horizon and AdS boundary are the same as in the s-wave case. The expansion of field $\Psi_x$ near the horizon is
\begin{eqnarray}\label{phorizon}
\Psi_x(r)&=& \Psi_{x0} + \mathcal{O}(r-r_h).
\end{eqnarray}
Again, because the equation of motion for $\Psi_x$ imposes a constraint on the derivative of $\Psi_x$ at the horizon, we have only one free parameter in the expansion (\ref{phorizon}). The expansion near the AdS boundary is
\begin{eqnarray}\label{pboundary}
\Psi_x(r)&=& \Psi_{xs} + \frac{\Psi_{xe}}{r^{d-2}}+... ~.
\end{eqnarray}
where $\Psi_{xs}$ is treated as the source of  the vector operator and $\Psi_{xe}$ is treated as the expectation value. Here we should note this expansion is not completely general. If d=2 \cite{Ren:2010ha,Jensen:2010em,Gao:2012yw}, a logarithmic term should be included.

Using the same strategy, we can also find the one parameter solution with $\Psi_{xs}=0$ numerically. Then the one parameter solution mimics the process of the spontaneous condensation of the p-wave operator. Here the equations of motion and the background metric are the same as in ref.~\cite{Gubser:2008wv}, thus the numerical results should be the same as those in ref.~\cite{Gubser:2008wv}.

\subsection{Scaling symmetry}
There is a scaling symmetry in the equations (\ref{eoms}), (\ref{eqpsix}) and (\ref{eqpsi3})
\begin{eqnarray}\label{scaling}
r\rightarrow \lambda r (r_h\rightarrow \lambda r_h),~ f\rightarrow \lambda^2 f,~ \phi\rightarrow \lambda \phi,~ \Psi_x\rightarrow \lambda \Psi_x ~ \Psi_3\rightarrow \Psi_3.
\end{eqnarray}
We can use this scaling symmetry to set the horizon radius to $r_h=1$. Then we can work in a fixed background metric. After we get the numerical solutions and some physical quantities, we can use again the scaling symmetry to recover the relevance with $r_h$, the latter is related to the temperature of the black brane.

The scaling behaviors of some physical quantities related to eq.~(\ref{scaling}) are
\begin{eqnarray}\label{Bscaling}
\mu \rightarrow \lambda \mu,~ \rho\rightarrow \lambda^{d-1} \rho,~ \Psi_{xs}\rightarrow \lambda \Psi_{xs}, \Psi_{xe}\rightarrow \lambda^{d-1} \Psi_{xe}~.
\nonumber \\
\Psi_{-}\rightarrow \lambda^{\Delta_-}\Psi_-,~ \Psi_{+}\rightarrow \lambda^{\Delta_+}\Psi_+.
\end{eqnarray}
Under this symmetry, the temperature scales as $T\rightarrow T \lambda$.  So we can build the invariant quantities under this scaling symmetry  such as
\begin{eqnarray}
\sqrt[d-1]{\rho}/T ,~ \sqrt[d-1]{\Psi_{xe}}/T,~
\sqrt[\Delta_-]{\Psi_{-}}/T,~ \sqrt[\Delta_+]{\Psi_{+}}/T.
\end{eqnarray}
These quantities are also dimensionless. We will use such dimensionless quantities in our numerical results.

\subsection{Comparing the s-wave and p-wave condensates}

\begin{figure}
\includegraphics[width=8cm] {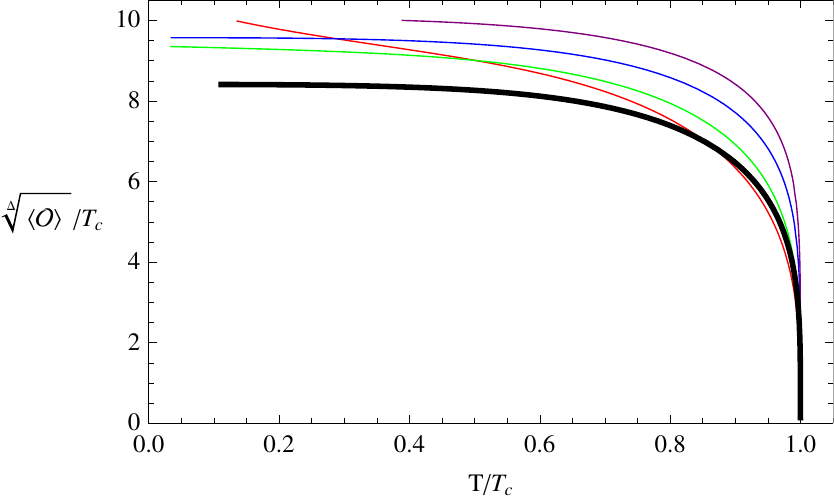}
\includegraphics[width=7.5cm] {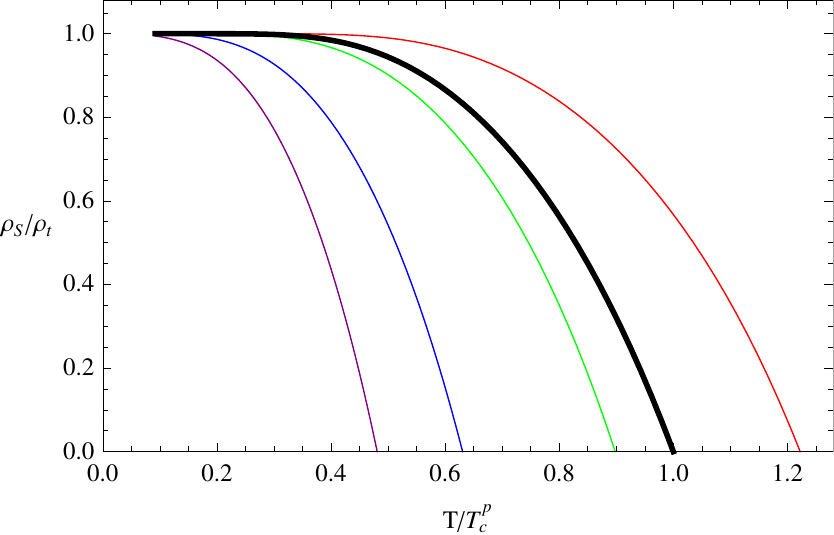}
\caption{\label{SorPCondensationRho} (Left) The condensation of operators. (Right) The ratio of the superconducting charge density over the total charge density. The thick black curve is for the p-wave phase and the thin colored lines are for the s-wave phase with operator dimension $\Delta=\frac{5}{3}(\text{red}), 2(\text{green}), \frac{5}{2}(\text{blue})$, and $3(\text{purple})$, respectively.}
\end{figure}

We see that both the s-wave and p-wave superconductivity can be realized in this model. In this subsection we present some numerical results.  We will work in $d=3$ in the rest of this paper. The condensate behavior of both the s-wave and p-wave orders is shown in the left plot in figure~\ref{SorPCondensationRho}, from which we can see that the condensate behavior of the p-wave operator is similar to the s-wave one, they both have a critical exponent $1/2$,  which is the typical value of mean field theory~\cite{Hartnoll:2008vx,Gubser:2008wv}.

But in order to compare the two phases with different condensates, it is necessary to find a same physical quantity that exists in the two phases. Because the s-wave operator and the p-wave operator are both coupled to the same U(1) gauge field in our model, we can choose a quantity related to field $\phi(r)$. One intersting quantity is the ratio of superconducting charge density over the total charge density.

The superconducting charge density equals the total charge density minus the normal charge density. The normal charge density is related to the charge carried by the black brane, and it is given by the horizon value of the electric field $\rho_n=\phi_1$, while the total charge density is related to the charge carried by the whole spacetime, and it is given by the electric field $\rho_t=(d-2)\rho$ at the AdS boundary.  Namely, the superconducting charge is carried by the scalar or vector hair between the horizon and the AdS boundary, and is given by
\begin{eqnarray}
\rho_s=\rho_T-\rho_n=(d-2)\rho-\phi_1.
\end{eqnarray}
We plot the ratio of $\rho_s/\rho_t$ for both the p-wave and s-wave phases in the right panel in figure~\ref{SorPCondensationRho}.  The ratio $\rho_s/\rho_t$ for the two condensate cases is also similar, they both have a critical exponent $1$ at the transition point. From the figure we can read off the critical temperature for all these condensations.  We see that the critical temperature $T_{c}^s$ of the s-wave condensate decreases as the dimension for the scalar operator $\Delta$ increases, but the critical temperature $T_{c}^p$ of the p-wave condensate is fixed. Our numerical calculation shows that the critical temperature of the p-wave condensate is equal to the one for the s-wave condensate with dimension $\Delta=\Delta_{cI}=1.8745$ in our model.

\section{Competition between the s-wave and p-wave condensates}\label{sect:competition}

We have realized both the s-wave and the p-wave superconductivity in our model. Thus we can study the competition between the two orders.
 From the previous section, we see that when one lowers the temperature of the strongly coupled system dual to this model, the phase with a higher critical temperature will occur first. If the dimension of
 the scalar operator $\Delta<\Delta_{cI}$, the s-wave condensate will occur first, otherwise the p-wave condensate happens first.

Once one condensate appears, if one goes on lowering the temperature of the system till the critical temperature of the other order, an interesting question arises: whether the other condensate happens or not.
To answer this, we should compare free energy of the system with different condensates. A phase with lower free energy is more thermodynamical favored.

We will work in the grand canonical ensemble with fixed chemical potential and compare the Gibbs free energy  $\Omega$ for the s-wave, p-wave and normal phases. In the AdS/CFT dictionary, the Gibbs free energy of the system is identified with the temperature times the Euclidean on-shell action of the bulk solution. In our case, because we work in the probe limit,  we need only calculate the contribution of matter fields to the
free energy in the different phases:
\begin{equation}
\Omega_m=T S_{ME},
\end{equation}
where $S_{ME}$  denotes the Euclidean action of matter fields in the black brane background. Note that since we work in the grand canonical ensemble and choose the scalar operator with dimension $\Delta_+$,  no additional surface term and counter term for the matter fields are needed. It turns out that the Gibbs free energy can be expressed as
\begin{equation}
\Omega_m=\frac{V_2}{g_c^2} (-\frac{1}{2}\mu\rho + \int_{rh}^\infty (\frac{\phi^2\Psi_x^2}{2f}+\frac{r^2 \phi^2\Psi_3^2}{f})dr).
\end{equation}
Here $V_2$ denotes the area of the 2-dimensional transverse space.

\begin{figure}
\includegraphics[width=7.5cm] {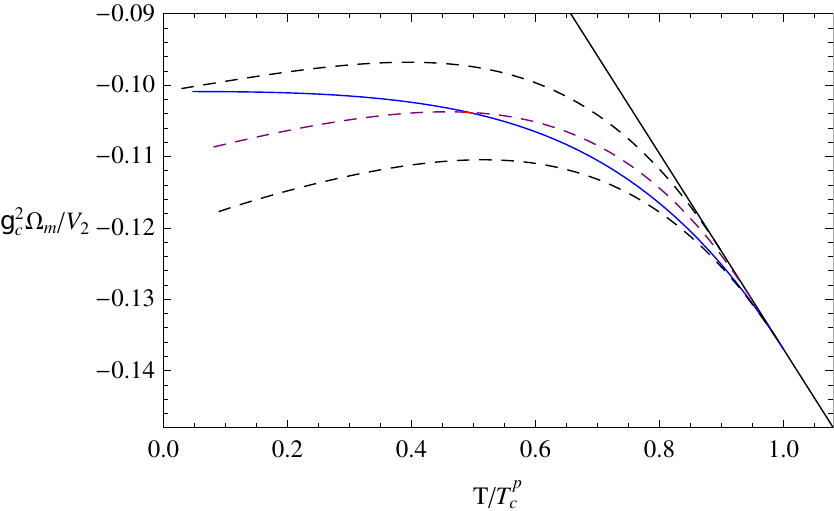}
\includegraphics[width=8.1cm] {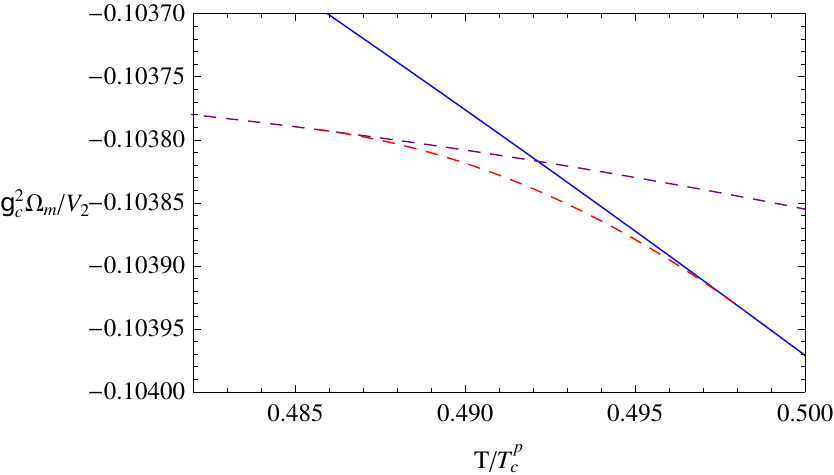}
\caption{\label{FreeEnergy} (Left) The Gibbs free energy versus temperature for various phases.  The black solid  curve is for the normal phase, the blue solid curve is for the p-wave phase, and the dashed lines from bottom to top are for the s-wave phase with operator dimension $\Delta=\Delta_{cI}$, $(6+\sqrt{3})/4$, and $\Delta_{cII}$, respectively. (Right) The Gibbs free energy in the region near the intersection point of the p-wave curve and the s-wave curve with $\Delta=(6+\sqrt{3})/4$. The dashed red curve is the free energy for the s+p coexisting phase with $\Delta=(6+\sqrt{3})/4$. The right figure is an enlarged version of the left one to show the s+p phase more clearly.}
\end{figure}

We plot the Gibbs free energy $\Omega_m$ for the different superconductivity phases in the left plot in figure~\ref{FreeEnergy}. There three typical (dashed) curves are shown for the s-wave phase with operator dimension $\Delta=\Delta_{cI}=1.8745$, $\Delta=\Delta_{eg}=(6+\sqrt{3})/4$(here the subscript $_{eg}$ stands for "e.g."="for example") and $\Delta=\Delta_{cII}=2.0000$, respectively.  The black solid curve represents the free energy for the normal phase without any condensation and the blue line is for the p-wave phase. Note that all the free energy curves for the s-wave and p-wave phases are tangent to the curve of the normal phase. This implies that both the s-wave and p-wave phase transitions are second order.

When $\Delta=\Delta_{cI}$, the s-wave condensate and p-wave condensate have the same critical temperature, and their values of the free energy are also the same at the transition point. But when one lowers the temperature, one can see that the free energy of the s-wave phase is always lower than the one of the p-wave phase.  Note the fact that the critical temperature increases as the scalar operator dimension decreases.
One then concludes that the free energy of the s-wave phases with $\Delta<\Delta_{cI}$ is always lower than that of the p-wave phase and the s-wave phase is always thermodynamically stable. In other words, in this case, once the s-wave condensate happens, the s-wave phase is always dominant and the p-wave condensate will never appear.

When $\Delta=\Delta_{cII}=2.0000$, the free energy curves for the s-wave phase and p-wave phase do not have any intersection until $T \approx 0$. So when the dimension of the scalar operator $\Delta>\Delta_{cII}$, the free energy of p-wave phase is always lower than the ones for the s-wave phase. Thus in that region, the p-wave phase is more thermodynamically favored.

The interesting case is the case with $\Delta_{cI}<\Delta<\Delta_{cII}$. In this case, the curves of the free energy for the s-wave phase and p-wave phase have an intersection when $T <T_c^p$. Take the case $\Delta_{eg}=(6+\sqrt{3})/4$ with $m^2=-33/16$ as an example, we can see that the free energy for the p-wave phase is lower when $T>T_{c}^{inter}$, while the free energy for the s-wave phase is lower when $T<T_c^{inter}$. When $T=T_c^{inter}$, they are equal.  This means that the system would be dominated by the p-wave phase as  $T>T_{c}^{inter}$ and be dominated by the s-wave phase as $T<T_{c}^{inter}$. At the point $T=T_{c}^{inter}$, a first order phase transition might occur if there do not exist other possible phases. But the story is more interesting than that. It turns out that in this region, there exists a so-called
s+p coexisting phase where both the s-wave condensate and p-wave condensate can happen at the same time. In the next section we will study the coexisting phase in some detail.

\section{The s+p coexisting phase and phase diagram}\label{sect:s+p}

To find the s+p coexisting phase, we need to solve the three equations (\ref{eqphi}),(\ref{eqpsix}) and (\ref{eqpsi3}) with both the three fields $\phi$, $\Psi_x$, and $\Psi_3$ turned on. The boundary conditions at the horizon and AdS boundary for the three fields are the same as those in the s-wave and p-wave phases. From the field expansions near the horizon, there are three shooting parameters $\phi_1$, $\Psi_{x0}$ and $\Psi_0$, in this case. Then to get the proper solutions, we should shoot the two boundary conditions $\Psi_-=0$ and $\Psi_{xs}=0$ with the three parameters. Therefore the problem we need to solve is to find the one parameter solution to the equations (\ref{eqphi}), (\ref{eqpsix}) and (\ref{eqpsi3}) with boundary conditions $\Psi_-=0$ and $\Psi_{xs}=0$. Here we stress that the s-wave phase, the p-wave phase and the normal phase are also solutions to the equations, with one or more functions trivial.

We still take the case with $\Delta=\Delta_{eg}=(6+\sqrt{3})/4$ as an example. If the s+p coexisting phase exists, it would emerge from the p-wave phase.  The expectation value of the s-wave operator will emerge continuously from zero to a finite value when one lowers the temperature. At the critical temperature, the expectation value for s-wave operator can be infinitely small, and then the function $\Psi_3(r)$ can be treated as a small perturbation in the p-wave phase. We can solve eq.~(\ref{eqpsi3}) with boundary condition $\Psi_-=0$ on the p-wave superconducting background. This is equivalent to find the instability of the s-wave mode on the p-wave superconducting background. Since eq.~(\ref{eqpsi3}) is linear for $\Psi_3(r)$, thus $\Psi_0$ can be scaled arbitrarily and $\Psi_0$ is not a free parameter that would affect the qualitative feature of the solution. So we fix $\Psi_0=1$ and tune the temperature of the background p-wave phase instead. We may reach the boundary condition $\Psi_-=0$ satisfied at some temperature $T_c^{sp1}$. If this happens, this temperature is just the critical temperature where the s+p coexisting phase occurs. Here we need to check the node of the solution $\Psi_3(r)$ because it is expected that the solution with one or more nodes is unstable.

Using the above strategy, we can find the critical temperature $T_c^{sp1}$ for the s+p coexisting phase. Then we can build the s+p coexisting phase by solving the coupled equations for $\phi$, $\Psi_x$ and $\Psi_3$.  It turns out that the s+p coexisting phase indeed exists in the region $\Delta_{cI}<\Delta<\Delta_{cII}$. The s+p coexisting phase starts from a pure p-wave phase at $T_c^{sp1}$. When one continues to lower the temperature, the p-wave condensate will quickly goes to zero and only the s-wave condensate remains, and the system goes into a pure s-wave phase at $T_c^{sp2}$. Using the p-wave condensate as a probe on the s-wave superconducting background, one can determine the critical temperature $T_c^{sp2}$.

\begin{figure}
\includegraphics[width=12cm] {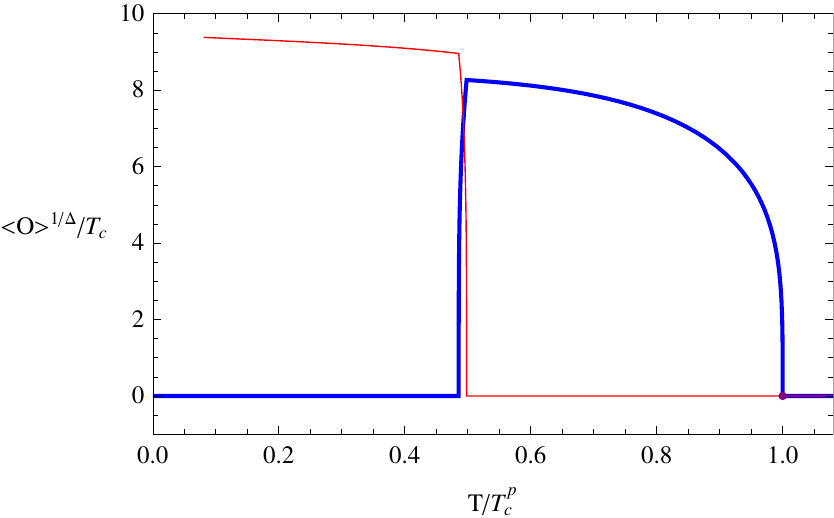}
\caption{\label{SPcondensation} Condensate of the operators in the s+p coexisting phase. The blue curve is for the condensate of the p-wave operator, while the red curve for the condensate of the s-wave operator.}
\end{figure}

For the case with $\Delta=\Delta_{eg}=(6+\sqrt{3})/4$, we find the two critical temperatures for the s+p coexisting phase are $T_c^{sp1}=0.4982T_c^p$ and $T_c^{sp2}=0.4856T_c^p$, respectively.   The s+p coexisting phase starts from the p-wave phase and ends with the pure s-wave condensate phase. We plot the condensate behavior for the coexisting phase in figure~\ref{SPcondensation}. We can see from the figure that when one lowers the temperature to $T_c^{sp1}$, the s-wave condensate emerges from a zero value. At the same time the s-wave condensate quickly repels the p-wave condensate so that the latter quickly goes to zero at temperature $T_c^{sp2}$. At this point, the s+p coexisting phase ends with a pure s-wave condensate phase.

The temperature region for the s+p wave coexisting phase is very narrow. This is similar to the situation of the coexisting phase with two s-wave orders in the probe limit studied in ref.~\cite{Basu:2010fa}.  However, for the latter, the region with the coexisting phase is enlarged with the full back reaction~\cite{Cai:2013wma}. This is due to the additional interaction between the two scalar fields in the bulk through gravity and  this interaction reduces the repellency between the two condensates. We expect that in our model, the
back reaction of matter fields will also enlarge the temperature region of the coexisting phase.

\begin{figure}
\includegraphics[width=12cm] {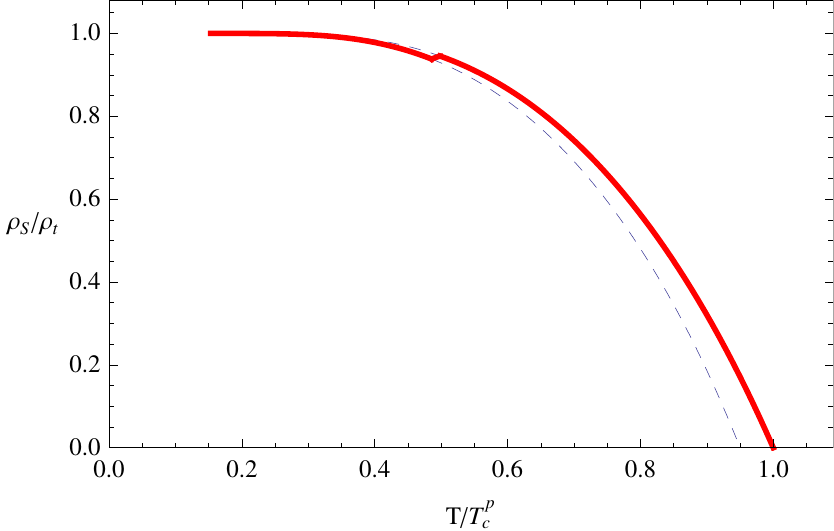}
\caption{\label{SPchargedensity} The ratio of the superconducting charge density over the total charge density $\rho_s/\rho_t$ versus the temperature $T/T_c^p$. The red curve describes the ratio $\rho_s/\rho_t$ when the system transfers from the p-wave phase to the s-wave phase through the s+p coexisting phase. The two dashed blue lines are for the ratio $\rho_s/\rho_t$ of the pure p-wave phase in temperature region $T<T_c^{sp1}$ and for that of the pure s-wave phase in temperature region $T>T_c^{sp2}$.}
\end{figure}

The superconducting charge density can be obtained from the $\phi$  field as the case in the pure s-wave or p-wave phase.  The ratio $\rho_s/\rho_t$  is plotted in figure~\ref{SPchargedensity}. We can see from figure that the ratio $\rho_s/\rho_t$ has a small kink in the region for the coexisting phase. When we lower the temperature, the ratio $\rho_s/\rho_t$ decreases in the s+p coexisting phase in contrast to the pure s-wave or p-wave phase. This might be an experimental signal of the phase transition from a single condensate phase to a coexisting phase, especially for the system with both s-wave and p-wave degrees of freedom as in ref.~\cite{Goswami:2013}.

It is also necessary to check the thermodynamical stability of this s+p coexisting phase. We calculate the Gibbs free energy $\Omega_m$ in the s+p coexisting phase and compare the result with those of the pure s-wave and p-wave phases in the right panel of figure~\ref{FreeEnergy}. We see that the s+p coexisting phase indeed has the lowest free energy and is thus thermodynamically favored in the temperature region. Thus the potential first order phase transition from the pure p-wave phase to the pure s-wave phase is replaced by the phase transitions from the p-wave phase to the s-wave phase trough an s+p coexisting phase. In other words, when we lower the temperature of the system, it first undergoes a phase transition from the normal phase to the pure p-wave phase at $T_c^p$. Then at $T_c^{sp1}$, a new phase transition occurs, and the system goes into an s+p coexisting phase.  At last the system undergoes the third phase transition from the s+p coexisting phase to a pure s-wave phase at $T_c^{sp2}$.  Note that all the three phase transitions are continuous ones, and are of characteristic of second order phase transition within the numerical accuracy.

\begin{figure}
\includegraphics[width=8cm] {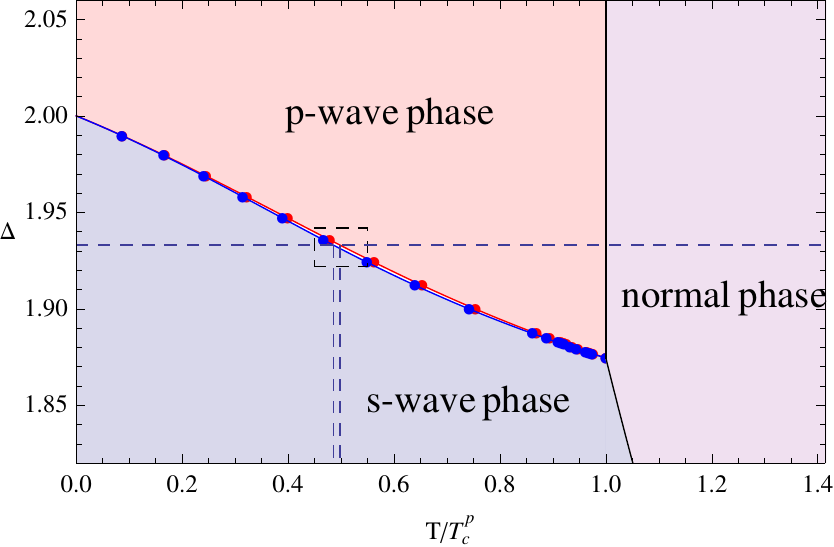}
\includegraphics[width=8cm] {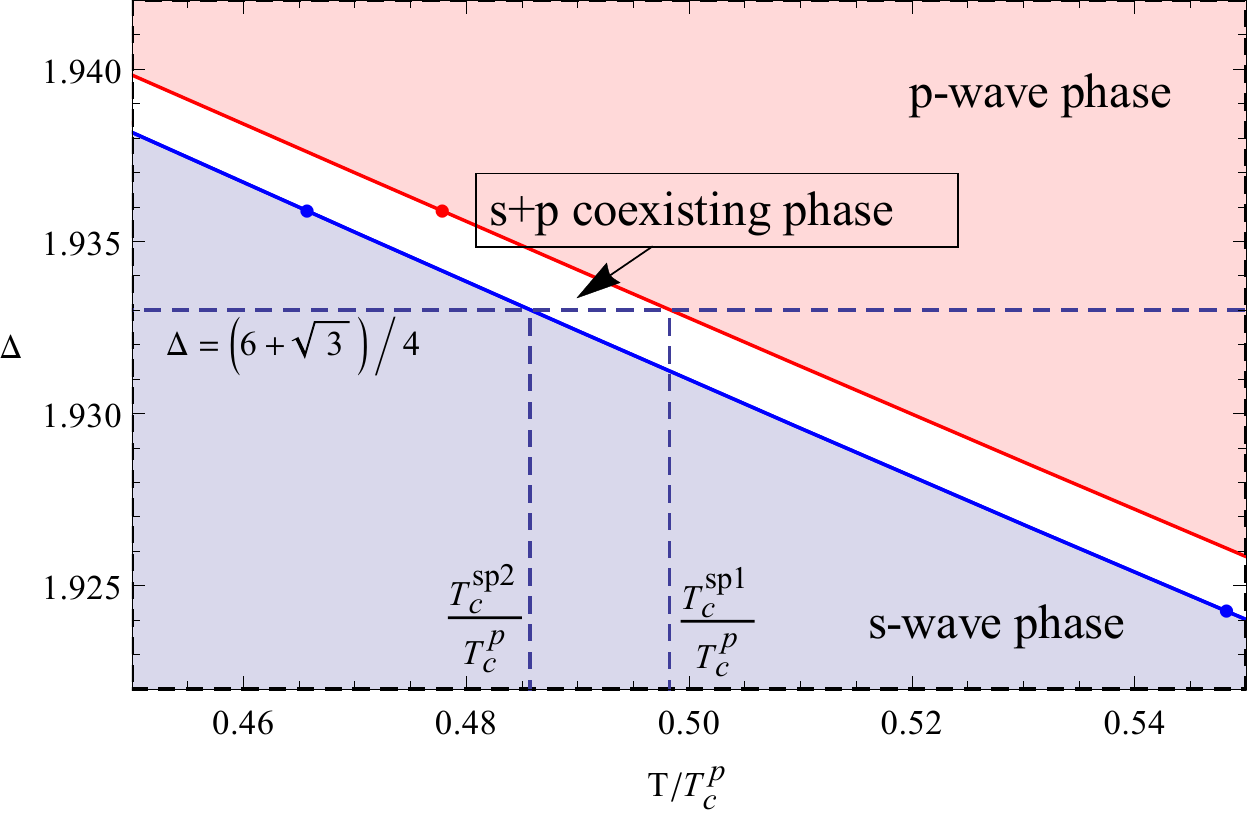}
\caption{\label{phasediagram}The $\Delta-T$ phase diagram. The right figure shows the dashed box area in the left figure. The three phases are colored differently. The black curve between the s-wave phase and the normal phase is made up by the critical points of the s-wave superconducting phase transition, while the vertical black line is determined by the critical temperature of the p-wave superconducting phase transition. The red curve describes the phase transition between the p-wave phase and the s+p coexisting phase and the blue one represents the phase transition between the s+p coexisting phase and the s-wave phase. The red and blue points are the data points from the numerical calculation, from which we get the red and blue curves respectively. }
\end{figure}

The Gibbs free energy curves of the s-wave and p-wave phases have an intersection when $\Delta_{cI}<\Delta<\Delta_{cII}$. The s+p coexisting phase just exists in this interval. We compute the values of $T_c^{sp1}$ and $T_c^{sp2}$ and get the relations $T_c^{sp1}(\Delta)$ and $T_c^{sp2}(\Delta)$ in the region $\Delta_{cI}<\Delta<\Delta_{cII}$ . With these data, we plot a phase diagram of our holographic model on the $\Delta$--$T$ plane in figure~\ref{phasediagram}. We can see from the figure that the system contains four kinds of phases known as the normal phase, the s-wave phase, the p-wave phase and the s+p coexisting phase. The normal phase dominates at high temperature, the s-wave phase dominates in the low temperature and small $\Delta$ area below the blue line, and the p-wave phase dominates in the low temperature and large $\Delta$ area above the red line. The s+p coexisting phase is favored in the area between the blue line and the red line. The region for the s+p coexisting phase is very narrow in the phase diagram, so we also plot an enlarged version in order to see this more clearly.

\section{\bf Conclusions and discussions}
\label{sect:conclusion}

In this paper, we set up a holographic superconductor model with a scalar triplet charged under an SU(2) gauge field in an AdS black brane background. In this model, the s-wave and p-wave condensates can be naturally realized.  The condensate behavior and the ratio of the superconducting charge density over the total charge density for the two phases are the same as in their individual model. We studied the competition between the two phases by comparing their critical temperature and free energy. It was found that the free energy curves of the s-wave and p-wave phases have an intersection when the dimension of the scalar operator satisfies $\Delta_{cI}<\Delta<\Delta{cII}$. In the same parameter region, the s+p coexisting phase exists with both the s-wave and the p-wave operators condensed. As a result, the holographic model admits totally
four phases, namely the normal phase without any condensate, s-wave phase, p-wave phase and the s+p coexisting phase. The latter three ones are superconductivity phases. We drawn the $\Delta$--$T$ phase diagram by numerical calculations. All phase transitions in the phase diagram are continuous in the probe limit. But we should note that once the back-reaction on the metric is included, the continuous phase transitions might change to first order ones as in ref.~\cite{Ammon:2009xh}.

We carefully investigated the s+p coexisting phase in the case with $\Delta=\Delta_{eg}=(6+\sqrt{3})/4$, by showing the condensate behavior of operators and the ratio $\rho_s/\rho_t$ of the superconducting charge density over the total charge density. The condensate picture shows that the s-wave and p-wave phases generally repel each other, but they can coexist in a narrow range of temperature. The slight decrease of the ratio $\rho_s/\rho_t$ might be a useful signal for detecting such an s+p coexisting phase experimentally.

Limited by the validity of the numerical calculations, we cannot obtain the data quite near the zero temperature limit. By extrapolation of our finite temperature data, we found that the two curves for the critical temperatures $T_c^{sp1}$ and $T_c^{sp2}$ of the s+p coexisting phase might intersect at $(T=0,\Delta=\Delta_{cII}=2.0000)$. Notice that the dimension of the p-wave operator is $\Delta_p=2$, it would be interesting to study the competition and coexistence of the two orders at zero temperature.

Note that the s+p coexisting phase in our model is quite similar to the s+s coexisting phase in the case with two s-wave operators coupled to the same U(1) gauge field~\cite{Basu:2010fa}.  It is found that with the full back reaction of the matter fields to the background geometry, the temperature region of the s+s coexisting phase is widely enlarged~\cite{Cai:2013wma}. This can be physically explained by the gravitational attractive interaction that reduces the repellency between the different condensates. Here we expect that the same story will appear in our model. The full back reaction of matter field will lead to a rich phase structure. This is left for our future work.

Note that in the present study, we only turned on a scalar component of the scalar triplet and two gauge
field components. It should be of great interest to investigate effects of other components in the matter sector in the holographic superconductor model and we believe some interesting features remain to be disclosed.

Finally let us mention here that we studied the holographic superconductor phase transition in the AdS
black brane background, it means that the dual field theory is at finite temperature. Note that a holographic superconductor/insulator phase transition model is already available by taking an AdS soliton
as the geometry background~\cite{taka}. Thus it is worthy to generalize the present study to the case of the superconductor/insulator phase transition.

\acknowledgments
ZYN would like to thank Lin Chen, Bin Hu, Ya-Peng Hu, Yun-Gui Gong, Li Li, George Siopsis, Yan-Qin Sui, Zhi-Yuan Xie, Bao-Chun Yang, Hai-Qing Zhang and Yun-Long
Zhang for helpful discussions. This work was supported in part by a grant from
State Key Laboratory of Theoretical Physics, Institute of Theoretical Physics, Chinese Academy of Sciences, in part by Shanghai Key Laboratory of Particle Physics and Cosmology under grant No.11DZ2230700, in part by the grants from Kunming University of Science and Technology and in part by the National Natural
Science Foundation of China under Grant Nos. 10821504 , 11035008 and 11247017.


\end{document}